\documentclass{aa}  
\usepackage{txfonts}
\usepackage{color}

\usepackage{graphicx}
\usepackage{txfonts}
\begin{document}

   \title{Stable Case BB/BC Mass Transfer to Form GW190425-like Massive Binary Neutron Star Mergers}
      \author{Ying Qin\inst{1,2}
          \and
          Jin-Ping Zhu\inst{3,4}
          \and
          Georges Meynet \inst{5,6}
          \and
         Bing Zhang\inst{7,8}
          \and
         Fa-Yin Wang\inst{9,10}
         \and
         Xin-Wen Shu\inst{1} 
          \and
         Han-Feng Song\inst{11,5}
          \and
         Yuan-Zhu Wang\inst{12}
           \and
         Liang Yuan\inst{1}
          \and
         Zhen-Han-Tao Wang\inst{13}
          \and
         Rui-Chong Hu\inst{7,8}
          \and
         Dong-Hong Wu\inst{1}  
          \and
         Shuang-Xi Yi\inst{14}  
          \and
         Qing-Wen Tang\inst{15} 
          \and
         Jun-Jie Wei\inst{2,16}           
          \and
         Xue-Feng Wu\inst{2,16}
          \and
         En-Wei Liang\inst{13}           
         }

   \institute{
            Department of Physics, Anhui Normal University, Wuhu, Anhui, 241002, China\\
              \email{yingqin2013@hotmail.com}
         \and
            Purple Mountain Observatory, Chinese Academy of Sciences, Nanjing 210023, China
        \and  
           School of Physics and Astronomy, Monash University, Clayton Victoria 3800, Australia
        \and
          OzGrav: The ARC Centre of Excellence for Gravitational Wave Discovery, Australia
        \and 
         Département d’Astronomie, Université de Genève, Chemin Pegasi 51, 1290 Versoix, Switzerland
        \and 
        Gravitational Wave Science Center (GWSC), Université de Genève, 24 quai E. Ansermet, 1211 Geneva, Switzerland
        \and 
            Nevada Center for Astrophysics, University of Nevada, Las Vegas, NV 89154, USA
        \and 
            Department of Physics and Astronomy, University of Nevada, Las Vegas, NV 89154, USA            
        \and      
            School of Astronomy and Space Science, Nanjing University, Nanjing 210093, People’s Republic of China 
         \and 
            Key Laboratory of Modern Astronomy and Astrophysics (Nanjing University), Ministry of Education, Nanjing 210093, People’s Republic of China
         \and
            College of Physics, Guizhou University, Guiyang, Guizhou 550025, PR China
         \and 
            Institute for Theoretical Physics and Cosmology, Zhejiang University of Technology, Hangzhou, 310032, China
         \and
            Guangxi Key Laboratory for Relativistic Astrophysics, School of Physical Science and Technology, Guangxi University, Nanning 530004, China
         \and
            School of Physics and Physical Engineering, Qufu Normal University, Qufu, Shandong 273165, China
         \and
            Department of Physics, School of Physics and Materials Science, Nanchang University, Nanchang 330031, China\\ \email{qwtang@ncu.edu.cn}
         \and 
            School of Astronomy and Space Sciences, University of Science and Technology of China, Hefei 230026, China}



 \abstract
 {On April 25th, 2019, the LIGO-Virgo Collaboration discovered a Gravitational-wave (GW) signal from a binary neutron star (BNS) merger, i.e., GW190425. Due to the inferred large total mass, the origin of GW190425 remains unclear.}
   {Assuming GW190425 originated from the standard isolated binary evolution channel, its immediate progenitor is considered to be a close binary system, consisting of a He-rich star and a NS just after the common envelope phase. We aim to study the formation of GW190425 in a solar-like environment by using the detailed binary evolution code \texttt{MESA}.}
   {We perform detailed stellar structure and binary evolution calculations that take into account mass-loss, internal differential rotation, and tidal interactions between a He-rich star and a NS companion. We explore the parameter space of the initial binary properties, including initial NS and He-rich masses and initial orbital period.}
   {We find that the immediate post-common-envelope progenitor system, consisting of a primary $\sim2.0\,M_\odot$ ($\sim1.7\,M_\odot$) NS and a secondary He-rich star with an initial mass of $\sim3.0-5.5\,M_\odot$ ($\sim5.5-6.0\,M_\odot$) in a close binary with an initial period of $\sim0.08-0.5\,{\rm{days}}$ ($\sim 0.08-0.4\,{\rm{days}}$), that experiences stable Case BB/BC mass transfer (MT) during binary evolution, can reproduce the formation of GW190425-like BNS events. Our studies reveal that the secondary He-rich star of the GW190425's progenitor before its core collapse can be efficiently spun up through tidal interaction, finally remaining as a NS with rotational energy even reaching $\sim10^{52}\,{\rm{erg}}$, which is always much higher than the neutrino-driven energy of the supernova (SN) explosion. If the newborn secondary NS is a magnetar, we expect that GW190425 can be the remnant of a magnetar-driven SN, e.g., a magnetar-driven ultra-stripped SN, a superluminous SN, or a broad-line Type Ic SN.}
   {Our results show that GW190425 could be formed through the isolated binary evolution, which involves a stable Case BB/BC MT just after the common envelope phase. On top of that, we show the He-rich star can be tidally spun up, potentially forming a spinning magnetized NS (magnetar) during the second SN explosion.}
  \maketitle
%
\section{Introduction}
On August 17th, 2017, the LIGO-Virgo Collaboration made the first observation of a Gravitational-wave (GW) signal from a binary neutron star (BNS) merger \cite[GW170817,][]{GW170817}. The source has component masses between $1.17-1.60\,M_\odot$ with a total mass of $2.74^{+0.04}_{-0.01}\,M_\odot$ considering the low-spin priors, in agreement with the masses of the known Galactic BNS systems. The second GW BNS merger (GW190425), which was detected in the third observing run (O3) of the LIGO-Virgo-KAGRA network \citep{gw190425} was found to have a total mass of $3.4^{+0.3}_{-0.1}\, M_\odot$, lying $5\sigma$ deviations away from the mean mass of Galactic BNS systems \citep{Farrow2019,Zhang2019}. To date, there is no confirmed detection of electromagnetic counterparts partially due to its large distance and poor sky localization \citep{Coughlin2019,Hosseinzadeh2019}. Furthermore, detailed investigations in \cite{Zhao2023} showed that heavier BNS systems are expected to generate kilonovae with lower peak luminosities and more rapid decay when compared to BNS systems with smaller total mass. Given the incredibly heavy inferred mass, our current understanding of the origin of GW190425 is still controversial.

The dominant formation channel for BNS systems is considered to be isolated binary evolution \citep[e.g.,][]{Tauris2017}. In this scenario, the primary star, which is initially more massive, evolves faster to become a NS after the first supernova (SN) explosion. The secondary star expands significantly after the main-sequence phase and then initiates mass transfer (MT) via the first Lagrangian point ($L_1$) onto the companion NS star. During this phase, the system is more likely to undergo a dynamically unstable MT phase, which probably leads to the formation of a common envelope \cite[so-called CE phase,][]{Paczynski1976}. If the system survives the CE phase, it becomes the immediate progenitor of a BNS, consisting of a He-rich star and a NS in a close orbit. Subsequently, the close binary system could undergo different MT phases, e.g., Case BB/BC MT \footnote{Case BB MT initiates from a He-rich star onto its companion when the donor star is burning helium in its shell, whereas Case BC MT occurs when carbon is ignited in the stellar core} \citep{Habets1986,Tauris2015,jian21}, depending on the evolutionary phase of the He-rich star. The survival of the second SN explosion can lead to the formation of the BNS system, which could become prime search targets for ground-based GW detectors, such as LIGO \citep{LIGO2015}, Virgo \citep{Acernese2015}, and KAGRA \citep{Aso2013}. 

In the standard isolated binary evolution channel, a fast-merging channel (namely unstable Case BB MT channel) is found to be inefficient in generating massive BNS systems \citep{Safarzadeh2020}. In this scenario, \cite{Galaudage2021} further estimated that the fast-merging binaries constitute 8\% - 79\% of BNS at birth and have a delay time (from the birth to the death) of $\sim\,5-401$ Myr (90\% credibility). \cite{RomeroShaw2020} pointed out that the constraint on the eccentricity ($e\leqslant0.07$ at 10 Hz with 90\% confidence) of GW190425 cannot provide evidence for or against the unstable MT channel. Early on, it was found in \cite{Vigna2018} that Case BB MT from He-rich stars onto NSs is most likely dynamically stable, which is in agreement with earlier findings in \cite{Tauris2015}.
Recent investigations in \cite{Giacobbo2018} found that, in order to produce BNSs with total masses of $3.2-3.5\,M_\odot$ from isolated binaries, the metallicity should be low, i.e., $\sim$ 5\% - 10\% solar metallicity ($Z_\odot$). Studies with population synthesis analysis from \cite{Kruckow2020} show the formation of GW190425 with Milky Way-like metallicity is possible. Alternatively, assuming the first-born NS with a typical mass of $\sim 1.4\, M_\odot$ and allowing for super-Eddington accretion onto NSs, \cite{Zhang2023} employed detailed binary evolution to find that the parameter space required to reproduce GW190425-like events is very limited. Additionally, fallback can contribute significantly to the mass growth of the newly formed NS, which may help explain the formation of heavy BNS systems like GW190425 \citep{Alejandro2021}.

It is worth noting that the inferred component masses are dependent on the choice of spin priors \cite[see their Figure 3 in][]{gw190425}. \cite{Zhu2020} proposed a spin prior, extrapolated from radio pulsar observations of Galactic binary NSs, finding positive support for a spinning recycled NS in GW190425. The inferred NS masses are $1.64_{-0.11}^{+0.13}\, M_\odot$ for recycled NS and $1.66_{-0.12}^{+0.12}\, M_\odot$ for slow NS, respectively. There is a slight discrepancy for inferred masses due to different spin priors, while the two masses are still consistent with the individual binary components being NSs. Additionally, the further constraint on the spin tilt angle \citep[$\lesssim 60^\circ$,][]{Zhu2020} shows that this system is consistent with the standard formation channel of isolated binary evolution. \cite{Mandel2021} proposed a probabilistic prescription for compact remnant masses, consistent with the formation of GW190425. 

In the context of a massive, binary evolution scenario for the formation of GW190425, the immediate progenitor is a close binary system of a He-rich star and a NS. In order to form a NS rather than a black hole, the mass of its progenitor He-rich star at solar metallicity needs to be less massive \cite[e.g., $\sim3.0-7.5\, M_\odot$,][]{Zhang2023}. After the core-helium burning phase, the star expands and then initiates MT onto the first-born NS through Case BB/BC MT. The late MT can further stripe He-rich star, producing Type Ib/c SNe \citep[e.g.,][]{Tauris2015}. If the system remains bound, the progenitor of the second-born NS could be significantly spun up via tidal interaction, leading to a fast-spinning magnetized NS \cite[namely magnetar, see details in][]{Hu2023}.

In this work, with the newly inferred component masses of NSs for GW190425, we perform detailed binary evolution with Modules for Experiments in Stellar Astrophysics (\texttt{MESA}) to explore its possible formation history. Therefore, the remainder of this paper is organized as follows. The main methods implemented in detailed binary evolution calculations are introduced in Section \ref{sect2}. In Section \ref{sect3}, we show the possible formation path of GW190425 and also estimate the rotational energy of the newly formed magnetar. Furthermore, we present in Section \ref{sect4} the helium envelope mass before the SN explosion and the associated ejecta mass. Finally, our main conclusions and some discussion are summarized in Section \ref{sect5}.

\section{Methods}\label{sect2}
We perform detailed binary modeling by using the release version \texttt{mesa-r15140} of the Modules for Experiments in Stellar Astrophysics (\texttt{MESA}) stellar evolution code \citep{Paxton2011,Paxton2013,Paxton2015,Paxton2018,Paxton2019,Jermyn2023}. We follow the same method as recent studies \cite[e.g.,][]{Hu2023,Fragos2023,lv2023,Zhang2023,Zhu2024gap} to create zero-age helium main sequence (namely ZamsHe) with different masses and adopt $Z_\odot$ = 0.0142 as the solar metallicity \citep{Asplund2009}.

We model convection using the mixing-length theory \citep{MLT1958} with a parameter of $\alpha_{\rm mlt} =$ 1.93. We adopt the Ledoux criterion to treat the boundaries of the convective zone and the step overshooting as an extension given by $\alpha_p = 0.1 H_p$, where $H_p$ is the pressure scale height at the Ledoux boundary limit. Semiconvection \citep{Langer1983} with an efficiency parameter $\alpha_{\sc}=1.0$ is adopted in our modeling. The network of \texttt{approx12.net} is chosen for nucleosynthesis. We treat rotational mixing and angular momentum transport as diffusive processes \citep{Heger2000}, including the effects of the Goldreich–Schubert–Fricke instability, Eddington–Sweet circulations, as well as secular and dynamical shear mixing. We adopt diffusive element mixing from these processes with an efficiency parameter of $f_c=1/30$ \citep{Chaboyer1992, Heger2000}. The $\mu$-gradient is sensitive to the rotationally induced mixing and here we reduce $\mu$-gradient by multiplying $f_\mu$ \cite[$f_\mu = 0.05$ suggested in][]{Heger2000}. 

Stellar winds of He-rich stars are modeled following the same method as in \cite{Hu2022}. We model the evolution of He-rich stars to reach central carbon depletion, from which the baryonic remnant mass is calculated following the ``delayed" supernova prescription \citep{Fryer2012}. For NSs, we convert baryonic to gravitational mass following the approach in \cite{Lattimer1989,Timmes1996}. We also consider the neutrino loss as in \cite{Zevin2020}. The maximum mass of a NS is assumed to be 2.5\,$M_{\odot}$. We calculate the timescale for orbital synchronization following \cite{Hurley2002} for massive
stars with radiative envelopes and adopt the updated fitting formula for the tidal coefficient $E_2$ provided in \cite{Qin2018}. We assume the Eddington-limited accretion onto NSs using the standard formulae \citep[see equation (4) in ][]{Tauris2017}. MT is modeled following the Kolb scheme \citep{Kolb1990} and the implicit MT method \citep{Paxton2015} is adopted.

\section{Formation of GW190425-like events and newborn magnetars} \label{sect3}
\begin{figure}[h]
     \centering
     \includegraphics[width=\columnwidth]{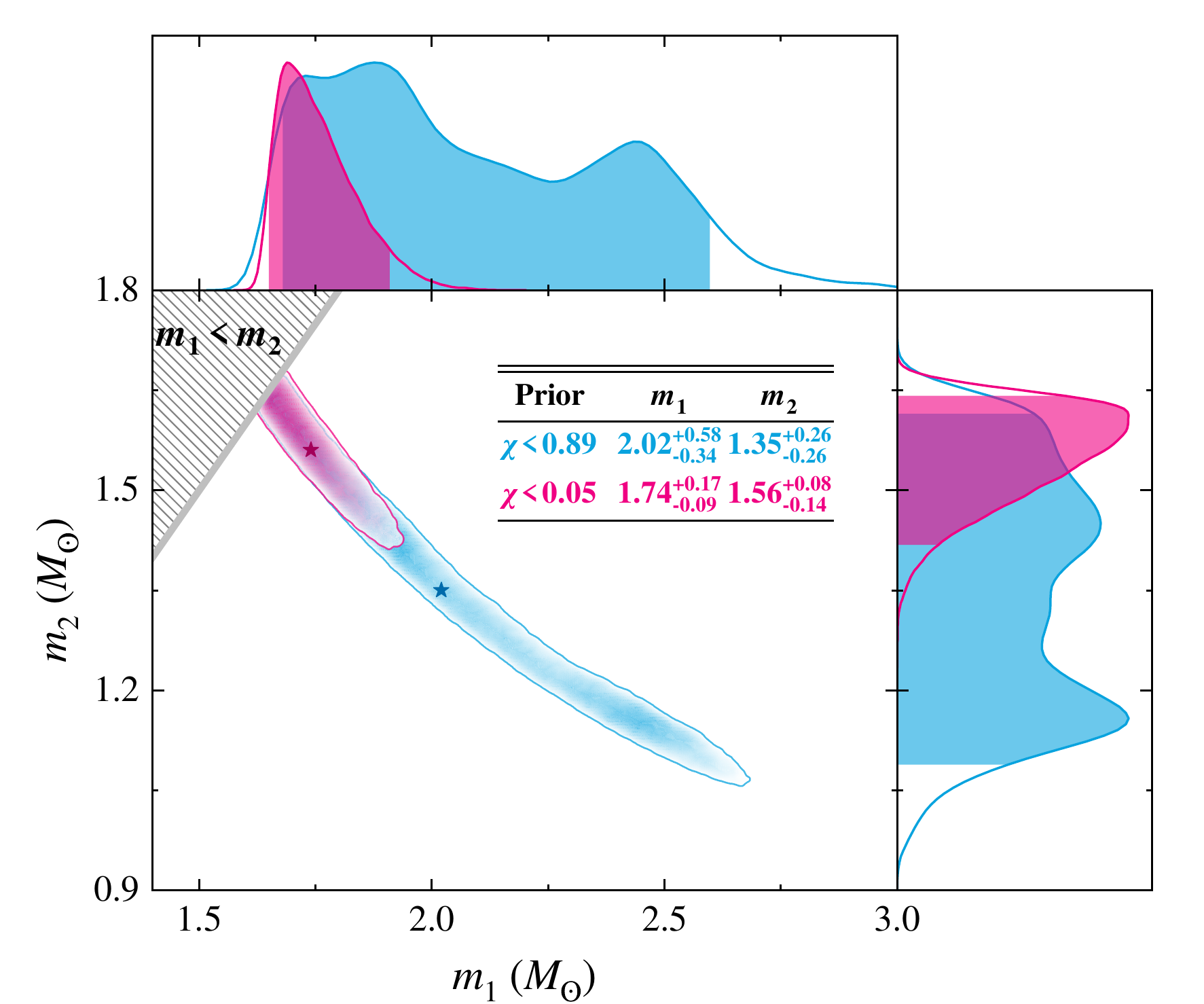}
     \caption{Posterior distributions of the primary and secondary masses of GW190425 by considering low-spin prior ($\chi<0.05$; pink) and high-spin prior ($\chi<0.89$; blue). The probability density functions of $m_1$ and $m_2$ normalized to have equal maxima are displayed in the top and right panels, respectively.} 
     \label{fig1}
\end{figure}

Hereafter, we directly use the newest release version\footnote{\url{https://gwosc.org/eventapi/html/GWTC-2/GW190425/v2/}.} of GW190425 within GWTC-2 to explore its formation pathway. We first employ the posterior data obtained from the \texttt{PhenomPNRT-LS} (\texttt{PhenomPNRT-HS}) template for the low-spin (high-spin) prior\footnote{We refer the dimensionless component spin magnitudes $<$ 0.05 ($<$ 0.89) to low-spin (high-spin) prior.}. Considering $90\%$ confidence intervals, the inferred primary and secondary NS mass are $m_1=2.02^{+0.58}_{-0.34}\, M_\odot$ and $m_2=1.35^{+0.26}_{-0.26}\, M_\odot$ ($m_1=1.74^{+0.17}_{-0.09}\, M_\odot$ and $m_2=1.56^{+0.08}_{-0.14}\, M_\odot$) using the low-spin (high-spin) prior. Figure \ref{fig1} illustrates the posterior distributions of the primary and secondary masses for GW190425, which are nearly consistent with the results in the discovered report of this BNS merger \citep{gw190425}.

With the newly inferred component masses of NSs, we perform detailed binary evolution of He-rich stars and NS as a point mass in close orbits. We model He-rich stars with initial mass $M_{\rm ZamHe}$ linearly from 2.5\,$M_\odot$ to 8.0\,$M_\odot$ at a step of 0.5\,$M_\odot$. We cover the initial orbital period, ranging from $0.04\, {\rm days}$ to $40\, {\rm days}$ with a logarithmic spacing of $\Delta\log(P/{\rm days})\approx0.16\,{\rm dex}$. We use $Z = Z_\odot$ as the initial metallicity of He-rich stars.

\subsection{Mass accretion onto NS}
\begin{figure}[h]
     \centering
     \includegraphics[width=\columnwidth]{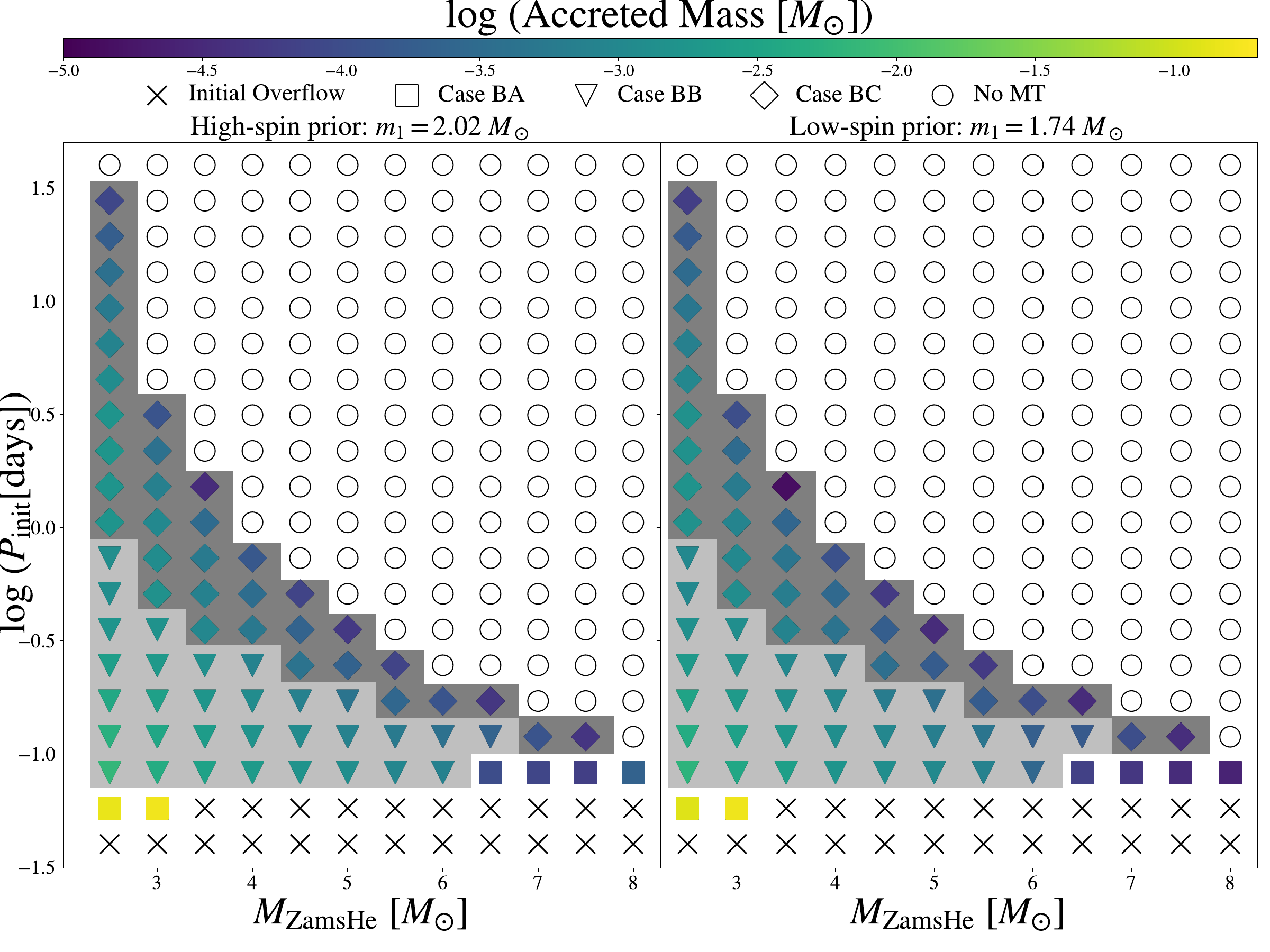}
     \caption{Accreted mass (the color bar) as a function of the initial orbital period and initial mass of He-rich stars. Cross: Initial Overflow; square: Case BA; triangle: Case BB; diamond: Case BC); circle: No mass transfer (MT). \textit{Left panel:} high-spin prior (NS mass $m_1 = 2.02\,M_\odot$), \textit{right panel:} low-spin prior (NS mass $m_1 = 1.74\,M_\odot$). We mark the parameter space of Case BB and Case BC with light and dark grey backgrounds, respectively.} 
     \label{Acc1.0}
\end{figure}

He-rich stars, especially for initially less massive, expand by around two orders of magnitude after leaving their core-helium burning phase \citep[see their Figure 1 in][]{Zhang2023}. Therefore, MT between a NS and a He-rich star via Roche-lobe overflow is expected to occur in a close orbit. In Figure \ref{Acc1.0}, we present the mass accreted onto NSs under different initial conditions of NS--He-rich star binary systems.

In the left panel of Figure \ref{Acc1.0}, the NS is assumed to have a mass of $m_1=2.02\,M_\odot$ (i.e., high-spin prior). First, MT in all binary systems is found not to occur for initial orbital periods $P_{\rm orb, init} \gtrsim 40\,{\rm days}$. Additionally, a longer initial orbital period is required for binary systems to experience mass interaction via $L_1$ as a He-rich star is less massive. This finding has been demonstrated in earlier investigations \cite[e.g.,][]{Ivanova2003,Tauris2015,Zhang2023}, which is because less massive He-rich stars expand significantly when evolving to become giant stars \cite[e.g., see the findings in][]{Fragos2023,Zhang2023}. Third, He-rich stars with an initial mass from 2.5\,$M_\odot$ to 7.5\,$M_\odot$ initiate MT onto NS companions when the central carbon is ignited (namely so-called Case BC MT, see the dark grey region). Compared with Case BC MT, He-rich stars can have MT during the shell-helium burning phase (i.e., Case BB MT phase, see light grey region) when the stars are less massive and the orbital periods are shorter (i.e., $P_{\rm orb}$ $\lesssim1.0\, {\rm days}$). We note that several systems with initially shorter periods are found to initiate MT even earlier, i.e., during the core-helium burning phase (Case BA MT, see the square symbols). For initial orbital periods $P_{\rm orb}\lesssim0.06\,{\rm days}$, systems are more likely to merge with their companions due to the initial overflow of He-rich stars at ZamsHe. Additionally, the parameter space of the binary systems to experience MT is the same when a massive NS is assumed (see the right panel of Figure \ref{Acc1.0}). 

\begin{figure}[h]
     \centering
     \includegraphics[width=\columnwidth]{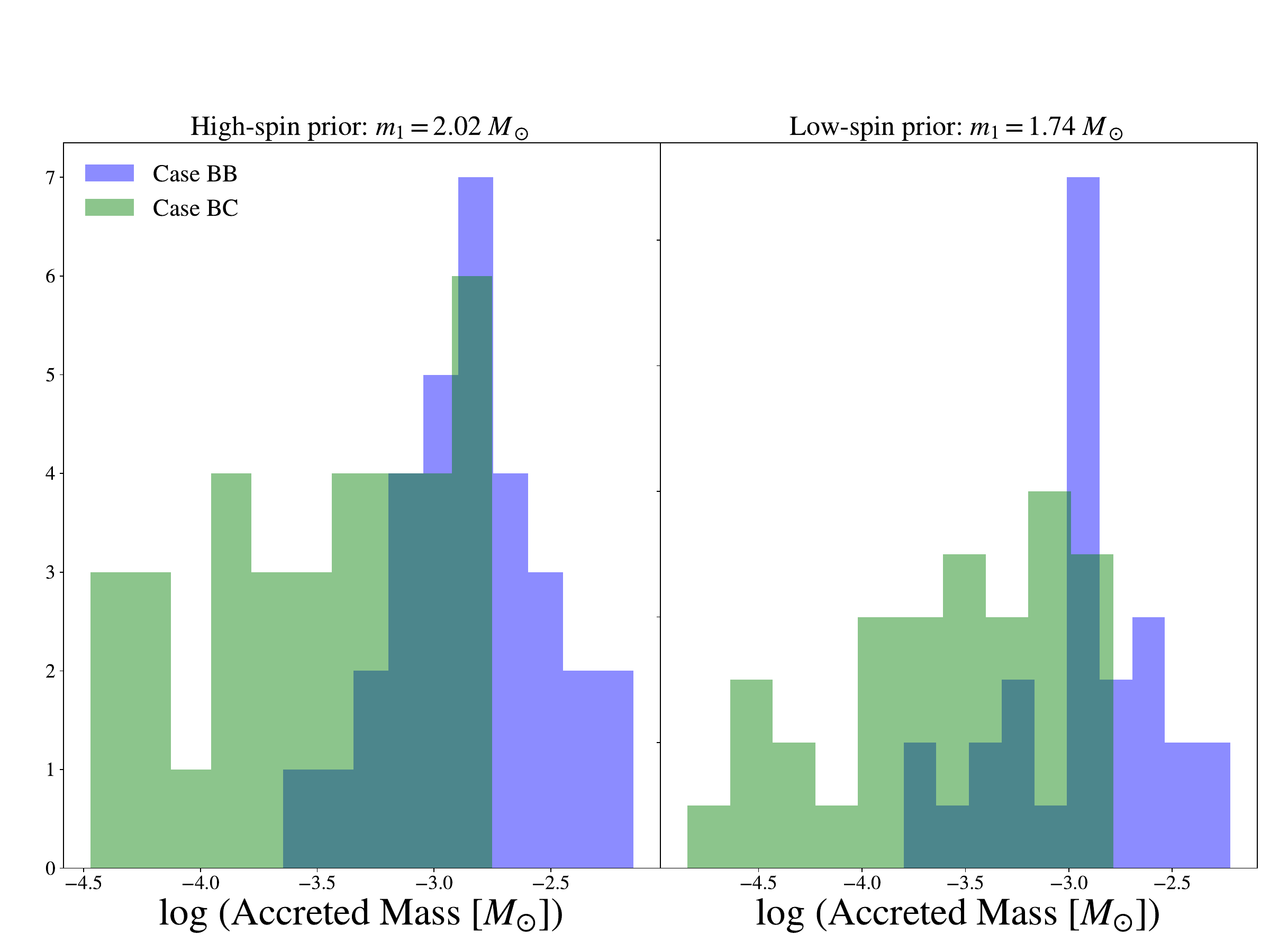}
     \caption{Histogram of accreted mass for Case BB (blue) and Case BC (green) MT. \textit{Left panel:} high-spin prior ($m_1 = 2.02\,M_\odot$), \textit{right panel:} low-spin prior ($m_1 = 1.74\,M_\odot$). All the He-rich stars have a metallicity of $Z = Z_\odot$.} 
     \label{Dist1.0}
\end{figure}

Figure \ref{Dist1.0} shows the accumulated mass accreted onto NSs through Case BB and Case BC MT, respectively. Given the NS companion mass of 2.02 $M_\odot$ (see the left panel), the accreted mass via Case BB MT is found to be a Gaussian-like distribution, in the range of $\sim 2.5 \times 10^{-4}-6.3 \times 10^{-3}\, M_\odot$ ($\log(M_{\rm accreted}/M_\odot) \in [\sim-3.6, \sim-2.2]$) and with a peak at $\sim 1.6 \times 10^{-3}\, M_\odot$ ($\log (M_{\rm accreted}/ M_\odot) \sim -2.8)$. In contrast, the accreted mass through Case BC MT is lower, approximately from $\sim 3.2 \times 10^{-5}\, M_\odot$ to $\sim 1.8 \times 10^{-3}\, M_\odot$ (log $(M_{\rm accreted}/M_\odot) \in [\sim -4.5, \sim -2.8]$). The difference in the accreted mass between Case BB and Case BC MT is ascribed to two factors. First, the MT rate of Case BB through the whole evolution is slightly higher when compared to Case BC MT. More importantly, the duration of the accretion for Case BB is much longer. These results are consistent with earlier findings in \cite{Zhang2023}. When considering a lower-mass NS companion (see the right panel), the corresponding mass accreted through Case BB and Case BC MT has a similar distribution, with the whole range slightly shifted to a lower end.

\subsection{GW190425 formed through stable mass transfer}
\begin{figure}[h]
     \centering
     \includegraphics[width=\columnwidth]{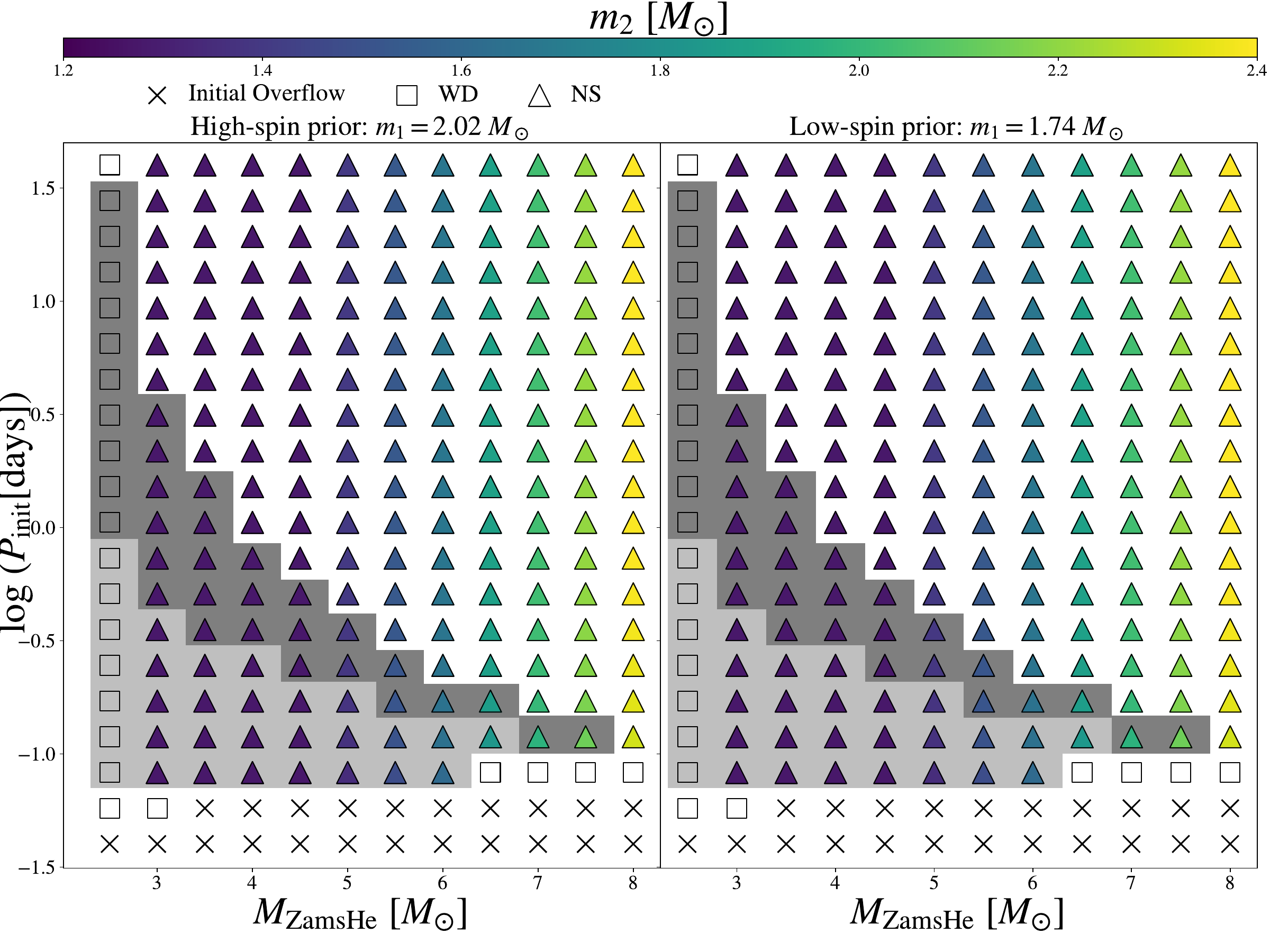}
     \caption{As in Figure \ref{Acc1.0}, but the color bar refers to the NS mass. Square: white dwarf; triangle up: NS formed through iron core-collapse SN (CCSN)}.   
     \label{m2_1.0}
\end{figure}

After the carbon is depleted in the center of He-rich stars, we adopt the ``\texttt{delayed}'' supernova prescription \citep{Fryer2012} to distinguish the genres of various compact objects (COs) and then calculate the baryonic remnant mass of NSs. In Figure \ref{m2_1.0}, we present the NS mass with different initial conditions of binary systems.
It is shown in the left panel that a white dwarf (WD) is formed when the initial He-rich star has a mass below $\sim 3.0\,M_\odot$, above which a NS is produced by the iron core-collapse supernova (CCSN). Notably, the mass of the newborn NS, determined by the inner structure of its progenitor before the SN stage, is independent of the initial orbital periods considered in this work and the companion star mass (see the right panel). He-rich stars at solar metallicity are expected to lose more mass due to metallicity-dependent winds \citep{vink2001}, producing NSs with mass ranging from $\sim 1.2\, M_{\odot}$ to $\sim 2.4\, M_{\odot}$ (see the color bar).

After the formation of double NSs, gravitational wave emission shrinks the orbits by removing their orbital angular momentum and leads to the merger of the COs. We adopt the following expression given in \cite{Peters1964} to calculate the merger time of BNSs, i.e.,
\begin{equation}
    T_{\rm merger}=\frac{5}{256} \frac{c^{5} a_{f}^{4}}{G^{3} (m_1 + m_2)^{2} m_{r}} T(e) , 
\end{equation}
where $c$ is the speed of light and $m_r$ is the binary’s reduced mass, $m_1$ and $m_2$ are the first-born and second-born NS, as well as $a_f$ the separation between the two components. For simplicity, we assume the orbit of BNS at its formation is circular (i.e., $T(e) = 1$). As we are mainly focused on BNS systems formed through stable MT after the post-CE phase, the SN kicks imparted onto the second-born NS are not taken into account. This has been pointed out in \cite{Tauris2015}
that the SN kicks of NSs formed from ultra-stripped progenitors are small (see more detailed discussions in Section 6.2). Similarly, electron-capture SN is generally expected to have small kicks \citep{Vigna2018,Giacobbo2019,Giacobbo2020}. For short orbital periods ($\lesssim$ 1 day), the system remains bound when a newly-born NS has a kick velocity of less than 100 km/s \citep[i.e., see their Figure 18 in][]{Tauris2017}.

\begin{figure}[h]
     \centering
     \includegraphics[width=1.0\columnwidth]{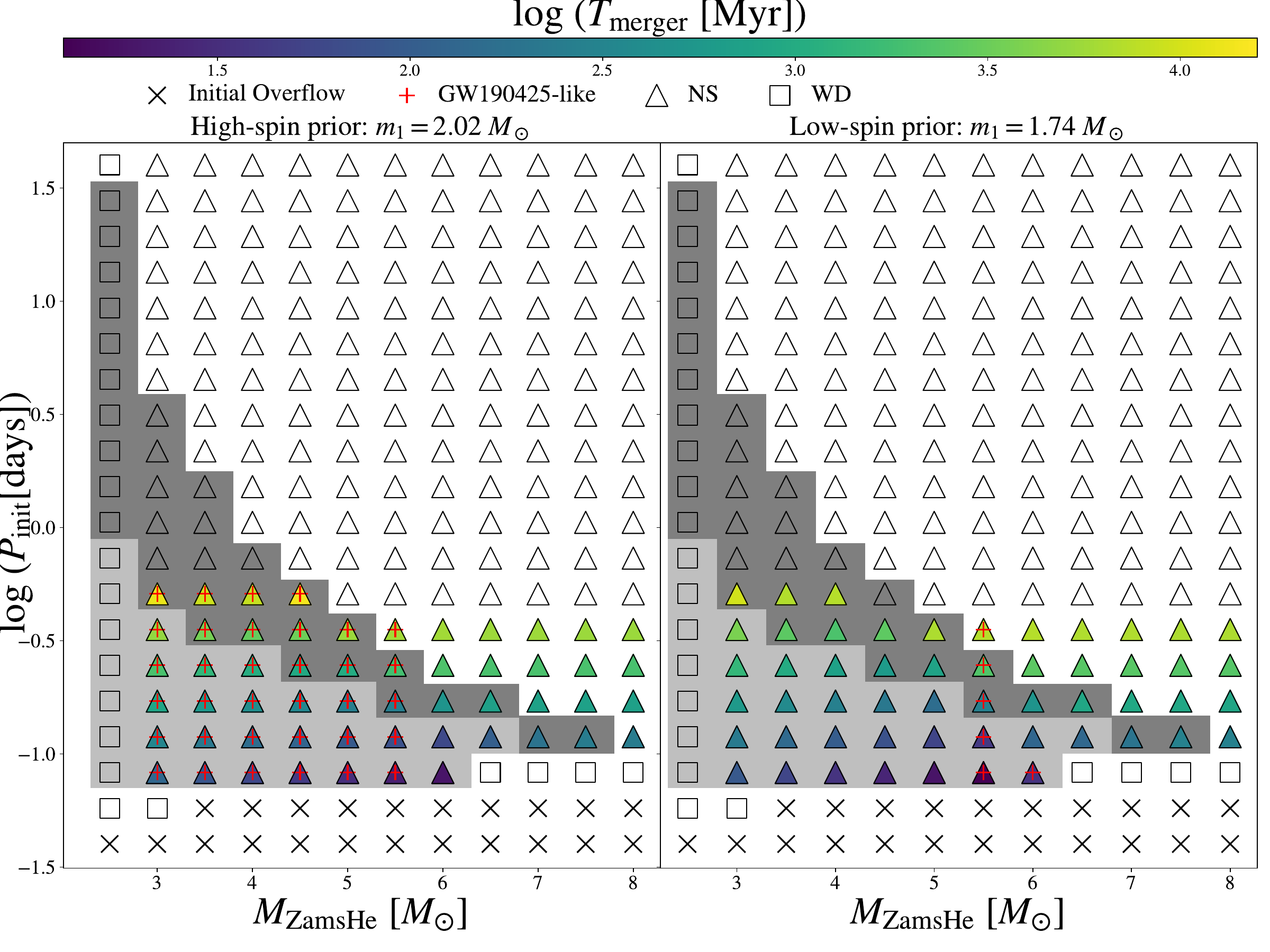}
     \caption{As in Figure \ref{Acc1.0}, but the color bar refers to the merger time of BNS due to the emission of GWs. We use the red plus symbol to mark the systems that resemble GW190425-like events.} 
     \label{tmer}
\end{figure} 

In Figure \ref{tmer}, we show the merger time ($T_{\rm merger}$, see the color bar) of BNSs with their $T_{\rm merger}\lesssim14\,{\rm Gyr}$ (the Hubble age). It is shown that GW190425-like events (see the red plus symbols) can be formed through either Case BC or Case BB MT depending on the specific initial conditions. In the left panel, in order to form observable BNSs (the colored symbols), the initial orbital periods need to be shorter than $\sim 0.5\,{\rm days}$ ($\log(P_{\rm init}/{\rm day})\sim -0.3$). Assuming the first-born NS mass of $m_2 = 2.02\, M_\odot$, we adopt the red plus symbols to mark the targets representing the GW190425-like events, with $\log (T_{\rm merger}/{\rm Myr}$) in the range of $\sim1.37-4.14$. We note that the parameter space to form GW190425-like events is rather limited, i.e., $3.0 \lesssim M_{\rm zamsHe}/M_\odot \lesssim 5.5$ and $0.08\lesssim P_{\rm init}/{\rm days} \lesssim0.5$. When considering the first-born NS mass of $m_1 = 1.74\, M_\odot$, the clear difference is that the immediate progenitor mass shifts to higher values, but with a rather narrow range, i.e., $5.5 \lesssim M_{\rm zamsHe}/M_\odot \lesssim6.0$. The merger time, however, is slightly shorter, i.e., $1.10\lesssim\log(T_{\rm merger}/{\rm Myr}) \lesssim3.85$.

\subsection{Rotational energy of the newly formed magnetar}
A NS accreting mass from its companion can be significantly spun up to become a mildly recycled pulsar \cite[see details in][]{Tauris2012,Tauris2015,Tauris2017}. The progenitor of the second-born NS if in close orbits, is expected to be synchronized by its companion through tides \citep{Detmers2008,Qin2018}, potentially resulting in fast-spinning highly-magnetized NS \citep{Hu2023}. 

\begin{figure}[h]
     \centering
     \includegraphics[width=\columnwidth]{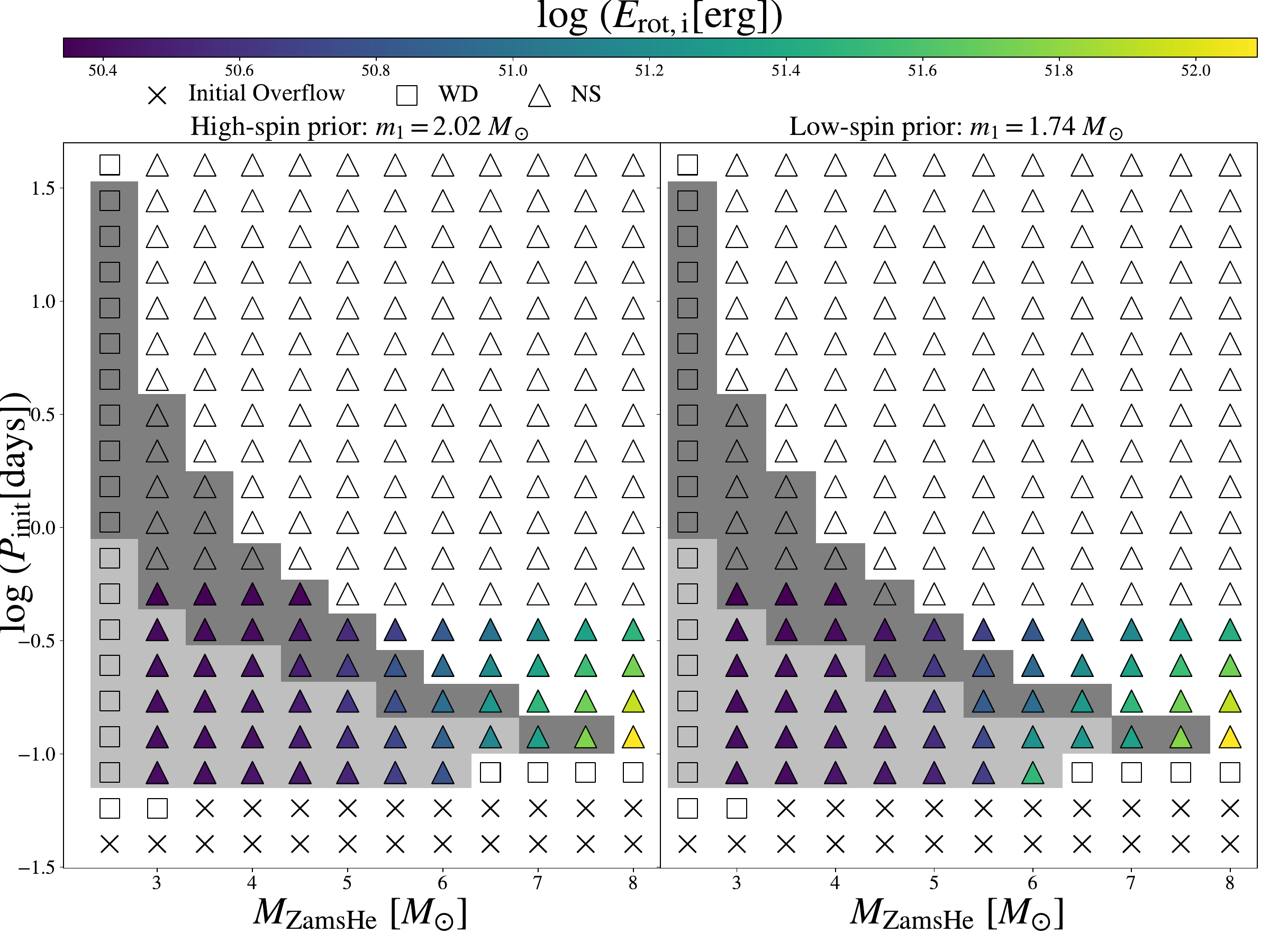}
     \caption{As in Figure \ref{Acc1.0}, but the color bar refers to the rotational energy of the newly formed magnetar in binary systems whose merger time is no longer than the Hubble age.} 
     \label{Etot}
\end{figure}

We calculate the magnetar's initial rotational energy, i.e., $E_{\rm rot, i} = 0.5 I_{\rm mag} (2 \pi / P_{\rm rot,i})^2$. In order to estimate the magnetar's moment of inertia, we adopt the empirical relation from \cite{Worley2008}, i.e.,

\begin{equation}
    I_{\rm mag} \approx 0.237 M R^2 \left [1 + 4.2\frac{M}{M_\odot}\frac{\rm km}{R} + 90\left (\frac{M}{M_\odot}\frac{\rm km}{R}\right )^4 \right ],
\end{equation}
where $M$ and $R$ are the magnetar's mass and radius, respectively. For the radius of a NS, we use a constant $R$ of 12.5 km \citep{gw190425,Landry2020}.

In Figure \ref{Etot}, we present the magnetar's initial rotational energy for the systems whose merger time is no longer than the Hubble age. It is worth noting that the magnetar's initial rotational energy varies in the range of $\sim 10^{50} - \sim 10^{52}$ erg. Therefore, the newly formed magnetar from tidal spin-up is a promising progenitor to produce stripped-envelope supernovae, including Type Ic superluminous supernovae, broad-line Type Ic SNe, and fast blue optical transients. More relevant studies can be found in \cite{Hu2023} who systematically investigated the origin of magnetar-driven stripped-envelope supernovae.

\section{Helium envelope mass before the SN explosion and its associated ejecta mass} \label{sect4}
\begin{figure}[h]
     \centering
     \includegraphics[width=\columnwidth]{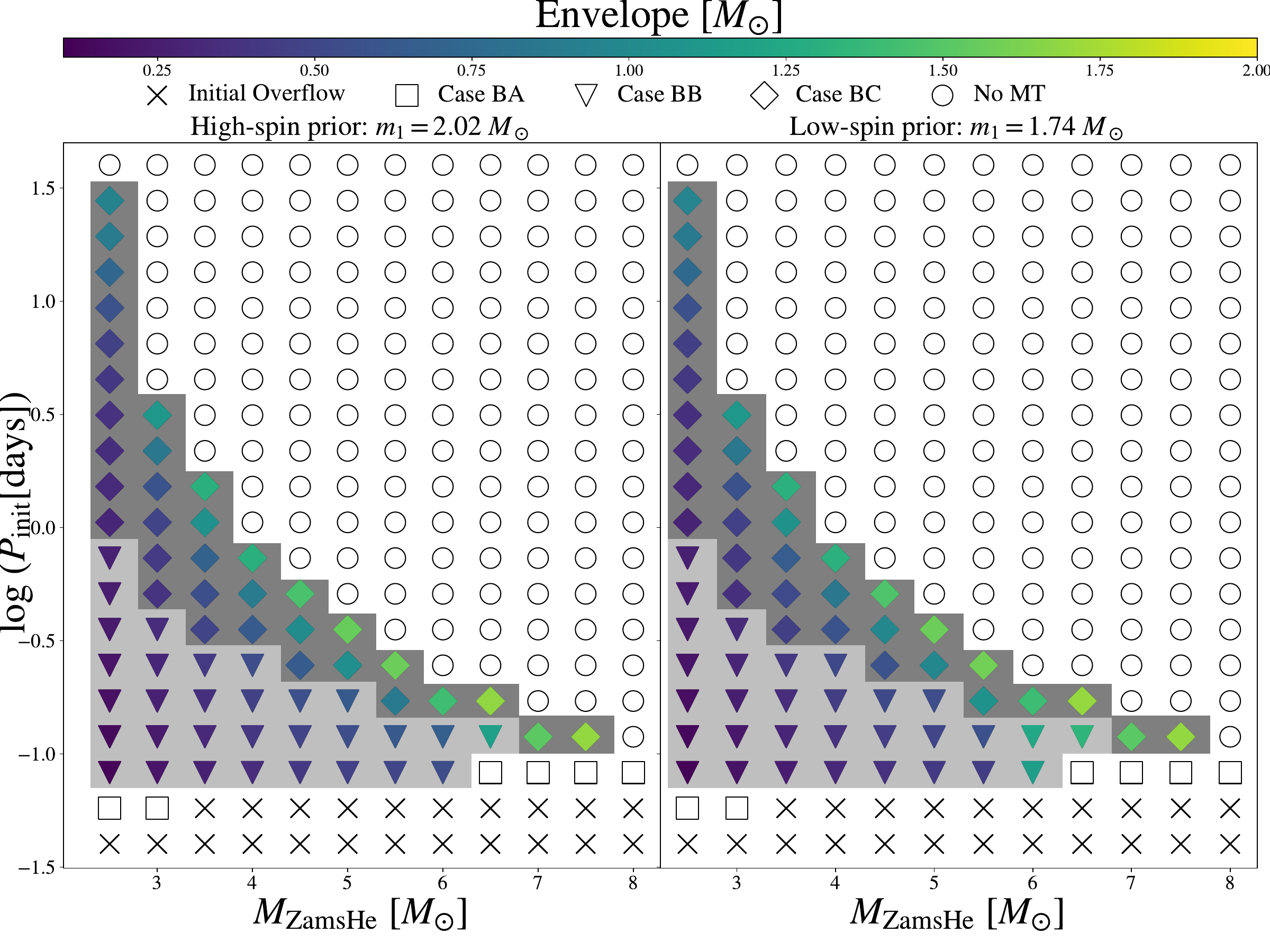}
     \caption{As in Figure \ref{Acc1.0}, but the color bar refers to the helium envelope mass (material outside the carbon-oxygen core) retained in He-rich stars before the SN explosion.} 
     \label{env}
\end{figure}

He-rich stars that experience Case BB/BC MT leading to further envelope loss are generally considered to produce ultra-stripped supernovae \citep[namely ultra-stripped SNe,][]{Tauris2015}. Figure \ref{env} shows the remaining envelope mass of He-rich stars before the SN explosion under different initial conditions. It is found that more envelope mass is retained in He-rich stars when increasing their initial orbital periods. This is because He-rich stars tend to be more significantly stripped in closer orbits. In the left panel (the companion NS mass $M_1 = 2.02\, M_\odot$), Case BC MT allows He-rich stars to retain helium envelope mass in the range of $\sim 0.3 - 1.7\, M_\odot$, while Case BB MT further strips He-rich stars, resulting in less helium envelope mass from $\sim 0.1\, M_\odot$ to $\sim 1.2\, M_\odot$. The envelope mass could be used as an upper limit of the ejecta mass during the SN explosion. As inferred in \cite{Hachinger2012}, the upper limit of helium mass required for type SNe Ic is $\sim 0.06 - 0.14\, M_\odot$. Therefore, our models indicate that stable Case BB MT in post-CE binaries can be considered a potential channel to contribute a certain proportion of all type Ic events \cite[i.e. $\lesssim 1\%$;][]{Tauris2013}. In contrast, more helium envelope mass could be retained for He-rich stars when experiencing Case BC MT, which is more likely to produce SNe Ib events. Therefore, GW190425-like sources are likely associated with type Ib/c SN events.

\begin{figure}[h]
     \centering
     \includegraphics[width=\columnwidth]{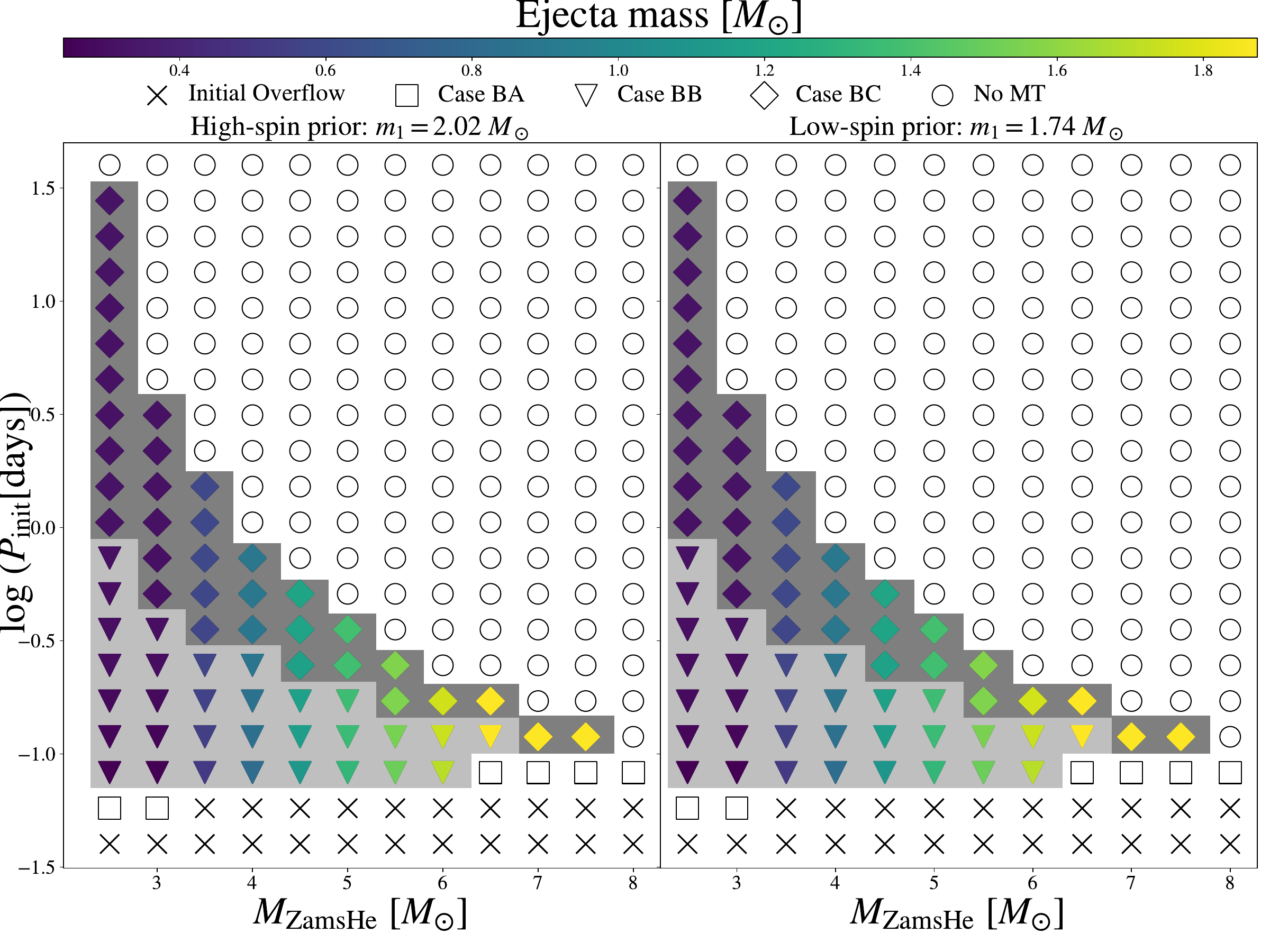}
     \caption{As in Figure \ref{Acc1.0}, but the color bar refers to ejecta mass.} 
     \label{ejecta}
\end{figure}

Recently, \cite{Hu2023} employed the population synthesis study to systematically investigate the evolution of He-rich stars with various companion stars, finding that He-rich stars in close orbits evolve to form fast-spinning magnetars. With the same method adopted in \cite{Hu2023}, we combine the carbon/oxygen core mass and the remnant mass to estimate He-rich stars/ ejecta mass (carbon/oxygen core mass - remnant mass). In Figure \ref{ejecta}, the ejecta mass is found to be in the range of $\sim 0.25 - 2.2\, M_\odot$. Therefore, the formation of GW190425-like events could be accompanied by some transients, e.g., type Ic superluminous supernovae, broad-line Type Ic SNe, and fast blue optical transients (FBOTs). We refer readers of interest to \cite{Hu2023} for more detailed calculations.

\section{Conclusions and discussion} \label{sect5}
We first obtain the component masses of GW190425, $m_1=2.02^{+0.58}_{-0.34}\,M_\odot$ and $m_2=1.35^{+0.26}_{-0.26}\,M_\odot$ ($m_1=1.74^{+0.17}_{-0.09}\,M_\odot$ and $m_2=1.56^{+0.08}_{-0.14}\,M_\odot$) at 90\% confidence intervals using the high-spin (low-spin) prior. Assuming the immediate progenitor of GW190425 is a He-rich star and a NS, we perf form a detailed binary evolution modeling to investigate its possible formation history and explore the properties of the relevant transients associated with the second SN explosion. In order to reproduce GW190425-like events with high-spin (low-spin) prior, the immediate progenitors should have an initial mass of a He-rich star in the range of $\sim 3.0- 5.5\, M_\odot$ ($\sim 5.5- 6.0\, M_\odot$) and an initial period from $\sim 0.08-0.5\,{\rm days}$ ($\sim 0.08- 0.4\,{\rm days}$), respectively. The mass accreted onto NS with mass inferred using the high-spin prior is found to vary from $\sim 2.5 \times 10^{-4}$ to $\sim 6.3 \times 10^{-3}\, M_\odot$ for Case BB MT and $\sim 3.2 \times 10^{-5}$ to $\sim 1.8 \times 10^{-3} M_\odot$ for Case BC MT, respectively. When considering a lower-mass (i.e., $1.74\, M_\odot$) NS companion, the corresponding mass accreted through Case BB and Case BC MT has a similar distribution, with the whole range slightly shifted to a lower end. Additionally, the corresponding merger time $\log(T_{\rm merger}/{\rm Myr}$) varies from $\sim$ 1.37 to $\sim$ 4.14 (from $\sim$1.10 to $\sim$ 3.85 for low-spin prior).

Due to subsequent mass transfer onto their companions, He-rich stars are further stripped and thus are more likely to produce ultra-stripped SNe \citep{Tauris2015}. We find that the remaining helium envelope mass before SN explosion for He-rich stars is in the range of $\sim$ 0.3 $\sim$ 1.7 $M_\odot$ for Case BC MT and $\sim 0.1- 1.2\, M_\odot$ for Case BB MT. Therefore, He-rich stars that experience extra mass transfer could potentially produce type Ib/c SN. We then estimate the ejecta mass in the range of $\sim 0.25- 2.2\, M_\odot$. Recently, \cite{Wu2022} found that less massive He-rich stars with mass of $\approx 2.5 -3\, M_\odot$ expand by a factor of a few during O/Ne-burning and thus can be further stripped due to later MT phase. Therefore, the ejecta mass estimated above is considered the upper limit.

We note that the SN 2023zaw, a sub-luminous and rapidly evolving SN with the lowest nickel mass, was recently reported to have a helium envelope mass of $\sim0.2\, M_{\odot}$ \citep{Das2024}. Their findings suggest an ultra-stripped SN, originating from a low-mass He-rich star in a close binary system. Furthermore, \cite{Moore2024} adopted Bayesian analysis to find that, in addition to the radioactive decay of $^{56} \rm Ni$, an extra energy source (e.g., a magnetar or interaction with circumstellar material) is required. As demonstrated in Section 3.3, He-rich stars in close orbits can be efficiently spun up \citep[e.g.,][]{Detmers2008,Qin2018,Sciarini2024}, forming fast-spinning magnetars. For the newly formed magnetar, we further estimate its rotational energy in the range of $\sim 10^{50} - 10^{52}$ erg, which is considered a promising progenitor to produce stripped-envelope supernovae, such as Type Ic superluminous supernovae, broad-line Type Ic SNe, and fast blue optical transients \citep[see detailed modeling in][]{Hu2023}. Recently, \cite{Siebert2024} reported a discovery of SN Ic, 2023adta identified in the deep James Webb Space Telescope (JWST)/NIRCam imaging. Follow-up observations with JWST/NIRSpec suggest the classification as abroad-line Type Ic SN. The estimated explosion parameters (i.e., ejecta mass and kinetic energy in their Table 5) for SN 2023adta are roughly consistent with the findings explored in this work. It should be noted that the spin period of the resultant NS is dependent on the magnetic braking \citep[e.g.,][]{Deng2021}, although it is unclear whether the He-rich star has a strong magnetic field on the surface.

It is possible that the formation scenario explored in this work could be applied to the formation of GW170817. On top of that, two high-confidence black hole-neutron star mergers \cite[GW200105 and GW200115,][]{Abbott2021BHNS} were found to have low misaligned spin, indicating that their immediate progenitors just after the CE phase consisted of a first-born black hole and low-mass He-rich star. The subsequent evolution very likely involves the stable Case BB/BC MT phase \citep{jian23}. More recently, the LVK collaboration \citep{GW230529} reported the observation of a coalescing compact binary (i.e., GW230529) with two component masses $2.5-4.5\, M_{\odot}$ and $1.2 - 2.0\, M_{\odot}$ (at the 90\% credible level). This binary system was inferred to have low effective inspiral spin, i.e., $\chi_{\rm eff} = -0.1_{-0.17}^{+0.12}$. We also note that GW190425 could be the merger of a light NS and a low-mass BH \citep{Han2020}. If the massive component is a black hole, their progenitor systems (GW230529 and GW190425) could share the same origin as GW200105 and GW200115. 

It is important to highlight that our findings could be dependent on the model adoption, e.g., remnant mass prescriptions \citep[e.g.,][]{Fryer2012}. Predictions of Galactic BNS mass distribution in \cite{Vigna2018} show that more massive components are expected (see their Figure 7) using the ``\texttt{delayed}'' supernova prescription when compared with ``\texttt{rapid}'' and ``\texttt{$\rm m \ddot{u} ller$}'' \citep{Muller2016} prescription. Therefore, the population of heavy BNS systems is expected to be small. However, predictions of the event rate for such systems are beyond the scope of this study. In the case of the high-spin prior, the primary NS mass of GW190425 was inferred to be $\sim2.0\,M_\odot$. This requires a massive NS to be born in the first SN explosion, e.g., PSR J1640+2224 \citep{Deng2020}, PSR J1614-2230 \citep{taur11}, and 2A 1822-371 \citep{wei23}. 

Notably, \cite{Moroianu2023NA} reported a possible association (2.8 $\sigma$ level) between GW190425 \citep{gw190425} and a fast radio burst (FRB 20190425A) \citep{FRB20190425}, which could be consistent with the FRB model invoking the collapse of a supermassive neutron star \citep{Falcke14} following a BNS merger \citep{Zhang14}. Further studies of \cite{Maga2024}, however, suggested that current state-of-the-art GW analyses disfavor the association between GW190425 and FRB 20190425A. Numerical simulations in \cite{Yamasaki2018} showed that a fraction of BNS mergers may form fast-rotating stable NSs, reproducing repeating FRBs \citep{CHIME2019} like FRB 121102. \cite{CHIME2019} proposed that a BNS merger channel could also form a millisecond magnetar \citep[also see][]{Wang2020}. For massive BNS systems, the primary component can be spun up through accretion but remains strong magnetic fields (Chu et al., in preparation), leading to a limited radio-pulsar lifetime and a low probability of being detected by radio emission \cite[also see claims in][]{Safarzadeh2020}. Their findings show that this could be the main reason that none of the observed Galactic BNSs is as massive as GW190425.

\begin{acknowledgements}
We thank the referee for constructive comments that helped improve the manuscript. We also thank Tassos Fragos and Wen-Cong Chen for their helpful comments on the manuscript. Y.Q. acknowledges support from the Anhui Provincial Natural Science Foundation (grant No. 2308085MA29), the National Natural Science Foundation of China (grant No. 12473036), and funding from the Key Laboratory for Relativistic Astrophysics at Guangxi University. J.P.Z. thanks the COMPAS group at Monash University. G.M. has received funding from the European Research Council (ERC) under the European Union's Horizon 2020 research and innovation program (grant agreement No 833925, project STAREX. Q.W.T acknowledges support from the Natural Science Foundation of Jiangxi Province of China (grant No. 20224ACB211001). This work was partially supported by the National Natural Science Foundation of China (grant Nos. 12065017, 12192220, 12192221, 12133003, 12203101, U2038106, 12103003). All figures are made with the free Python module Matplotlib \citep{Hunter2007}.
\end{acknowledgements}

\bibliographystyle{aa}
\bibliography{ref}
\end{document}